\documentclass[conference]{IEEEtran}

 \usepackage{cite,url}
\usepackage{amsfonts,amssymb,verbatim}
\usepackage{mathtools, cuted}
\usepackage{lipsum, color}
\usepackage{amsmath,amssymb,graphicx,bm,color,amsbsy,amsthm,algorithm,algorithmic,epsfig,subfigure,lipsum,multicol}
\usepackage{epstopdf}
\usepackage[makeroom]{cancel}

\begin{document}

\title{RF Source Seeking using Frequency Measurements}

\author{\IEEEauthorblockN{Muhammed Faruk Gencel\IEEEauthorrefmark{1}, Upamanyu Madhow\IEEEauthorrefmark{2}, Jo\~ao Pedro Hespanha\IEEEauthorrefmark{3}}
\IEEEauthorblockA{Department of Electrical and Computer Engineering\\
University of California Santa Barbara\\
Santa Barbara, California 93106\\
Email: \IEEEauthorrefmark{1}gencel@ece.ucsb.edu, \IEEEauthorrefmark{2}madhow@ece.ucsb.edu, \IEEEauthorrefmark{3}hespanha@ece.ucsb.edu}
}

\maketitle
\begin{abstract}
In this paper, we consider a problem motivated by search-and-rescue applications, where an unmanned aerial vehicle (UAV) seeks to approach the vicinity of a distant
quasi-stationary radio frequency (RF) emitter surrounded by local scatterers. The UAV employs only measurements of the Doppler frequency of the received RF signal, along with its own bearing, to continuously adapt its trajectory. We propose and evaluate a trajectory planning approach that addresses technical difficulties such as the unknown carrier frequency offset between the emitter and the UAV's receiver, the frequency drifts of the local oscillators over time, the direction ambiguity in Doppler, and the noise in the observations. For the initial trajectory, the UAV estimates the direction of the emitter using a circular motion, which resolves direction ambiguity. The trajectory is then continuously adapted using feedback from frequency measurements obtained by perturbing the bearing around the current trajectory.  We show that the proposed algorithm converges to the vicinity of the emitter, and illustrate its efficacy using simulations. 


\end{abstract}


\section{Introduction \label{sec:intro}}

We consider a scenario motivated by search-and-rescue, or other emergency applications, where a UAV seeks to approach an RF source, starting from an initially large distance.The UAV is equipped with a single omnidirectional antenna, and does not rely on GPS or on being able to decode messages from the emitter.  The source may be surrounded by local scatterers. In the approach proposed
here, the UAV adapts its trajectory towards the emitter using frequency measurements on the received beacon. In an ideal line of sight (LoS), a single omni-directional antenna can extract the angle of the arrival $\theta$ between the velocity vector of the mobile node and LoS to the source by measuring the Doppler frequency $f_d = \frac{v \cos \theta}{c} f_c$, where $v$ is the velocity and $c$ is the speed of light.  Thus, 
a natural approach is for the UAV to follow the trajectory that maximizes the Doppler shift (which corresponds to $\theta = 0$).
%
However, translating this intuition into a working approach requires that we address the following technical challenges:\\
1) The scattering environment around the source causes multipath fading, resulting in large spatial variations of the received signal power. This can often lead to errors in frequency measurements, especially at the low received signal-to-noise ratio (SNR) obtained at large distances.\\
2) The local oscillators at the emitter and UAV are not synchronized, and drift over time. Thus, the frequency measurements made by the UAV are a sum of the Doppler shift and a slowly drifting carrier frequency offset.\\
3) Even in ideal LoS settings, Doppler estimates have direction ambiguity: if the trajectory makes an angle $\theta$ with the LoS, then the Doppler shift is proportional to $\cos \theta$, and cannot therefore enable us to
distinguish between $+ \theta$ and $- \theta$.\\
4) Any trajectory adaptation done by the UAV should be feasible, avoiding sharp direction changes.

The main contribution of this paper is to show that we can indeed overcome the preceding difficulties to obtain a scheme that reaches the vicinity of the emitter, with net distance traversed being only a small fraction 
(of the order of 10\%) larger than the initial LoS distance between the UAV and the emitter. We consider a small UAV that flies at around 100 m altitude, listening to a beacon in commercial frequency bands (the carrier frequency is set to 2 GHz in our numerical examples). The initial distance between the UAV and the source is of the order of 5 km. In our proposed approach, the UAV obtains an initial trajectory estimate by finding the direction of maximum Doppler when executing a circular motion.  Subsequently, it employs feedback control to continuously adapt its trajectory, using the change in measured frequency offset as it executes designed piecewise linear deviations in bearing from the nominal trajectory.

The proposed approach performs significantly better than RF source following using received signal strength (RSS) measurements \cite{wadhwa2011following}.  
While RSS measurements are simpler to make and do not require coherent processing at the receiver, 
the sensitivity of RSS change as a function of range to the emitter is small even in ideal settings, since it is proportional to the inverse square of the range.  In addition, local scatterers around the emitter lead to slow, and deep, spatial variations in RSS due to fading.  The approach in \cite{wadhwa2011following} employs the observation that the rate of change of RSS due to fading is minimum
in the LoS direction, along with a random walk inspired by bacterial chemotaxis. For a setting similar to ours, the RSS-based scheme requires the UAV to traverse a distance 
that is about three times larger than the shortest path between the initial UAV location and the emitter.  Furthermore, the trajectories employed in \cite{wadhwa2011following}, both for
initialization and for the random walk, are non-smooth and difficult to execute.

RSS measurements are more effective if supplemented with {\it directional} information.  A rotating UAV was employed in \cite{venkateswaran2013rf}, with the angle of arrival to the emitter
estimated as the direction of maximum RSS.  This approach is not applicable to fixed wing UAVs with omnidirectional antennas as considered here.



While we consider the problem of {\it approaching} the emitter, there is a significant literature on {\it localizing} the emitter using a mobile 
platform \cite{becker1999passive, nguyen2016single} or multiple mobile platforms \cite{amar2008localization,tahat2016look}. Particle filter based algorithms for tracking the posterior distribution of the emitter are investigated in
\cite{witzgall2011Doppler,witzgall2015single}.  In addition to having a different design goal from ours, it is worth noting that these approaches
require that the mobile platform always knows its own absolute location, say using GPS.  Our problem formulation requires the UAV to track changes in its own bearing, but does not require that it know its absolute location, and hence is applicable even in GPS-denied environments.

\begin{figure}[htbp]
\centering
\includegraphics[width=1\columnwidth]{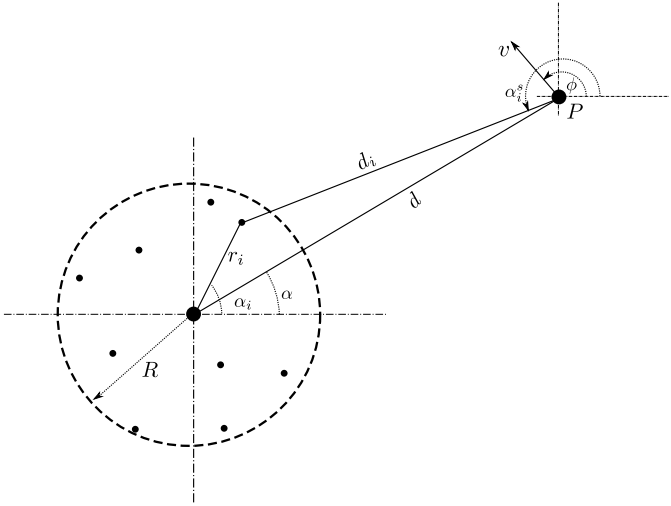}
\caption{2D system model, with scatterers in a disk around the source}
\label{fig:system2D}
\end{figure}

\section{System Model \label{sec:model}}

The problem of drawing the UAV to the emitter location is three-dimensional, but we restrict the problem to two dimensions  for simplicity. As shown in Figure \ref{fig:system2D}, the source is located at the the origin of the 2D plane, and is surrounded by $L$ local scatterers.  The UAV coordinates at any given time are denoted by $p = (x,y)$ with velocity components $v_{x} = v \cos(\phi)$ 
and $v_{y}= v \sin(\phi)$. The distance between the mobile receiver and the source is $d$. The scatterers are inside an annulus with outer radius $R$ and inner radius $R_{in}$, and the angles $\{ \alpha_i \}$ are uniformly distributed between $-\pi$ and $\pi$ for each scatterer.

We consider a narrowband flat fading channel at carrier frequency $f_c$, and assume that the receiver will compute its estimates based on 
known pilot signals transmitted by the source. The complex baseband channel seen by the mobile node can be expressed as the sum of LoS and scattered components and can be written as 
\begin{equation} \label{eq:channel}
\begin{split}
h(t) =& \sqrt{\frac{K \sigma_h^2}{K+1}} e^{j (2 \pi f_{d,max} \cos(\alpha-\phi) t + 2 \pi f_{o}(t) t+ \psi_0)} +  \\
& \sqrt{\frac{\sigma_h^2}{K+1}} \sum_{i=1}^{L} e^{j (2 \pi f_{d,max} \cos(\alpha^s_i-\phi) t + 2 \pi f_{o}(t) t+ \psi_i)}  + n(t)  
\end{split}
\end{equation}
where $K$ is the ratio of power between the direct path and the scattered paths,  $f_{d,max} = \tfrac{v f_c}{c}$ is the maximum Doppler frequency, $f_{o}(t)$ is the carrier frequency offset drifting over time, $\psi_0$ and $\psi_i$ are the phase of LoS and scattered signal components respectively, $\sigma_h^2$ is the received signal power and $n(t)$ represents the additive noise at the receiver with the variance $\sigma_n^2$. 

The received signal strength $\sigma_h^2$ is governed by the distance between source and mobile receiver $d$, along with spatial variations due to multipath fading. The effect of Doppler frequency on the received signal profile is negligible at these low speeds since $f_c \gg f_{d,max}$. We model the received signal strength by modeling the electric field at the mobile node at a point P in polar coordinates $(d,\alpha)$ as\cite{wadhwa2011following}: 
\begin{equation} \label{RSS}
EF(d,\alpha) = \frac{e^{-j \beta d}}{d} + \sum_{i = 1}^{L} \frac{ \Gamma_i e^{-j \beta (d_i+r_i)}} {d_i+r_i} 
\end{equation}
where $\beta = \tfrac{2 \pi}{\lambda}$ and $\lambda$ is the wavelength, $d_i+r_i$ is the total distance of the path that goes from the source to receiver through i$^{th}$ scatterer and $\Gamma_i$ is the reflection coefficient for the i$^{th}$ scatterer \cite{aragon2008antennas}.

Frequency estimation accuracy in a flat fading channel, assuming all paths have the same frequency offset, is proportional to the received SNR \cite{baronkin2001cramer}.
This assumption is a good approximation for our model when $d \gg R$.  We use low-complexity single tone frequency estimation \cite{brown2010low}, selecting the maximum frequency over the DFT grid, and then interpolating using a quadratic fit. As the UAV gets closer to the source, each path sees a different frequency offset due to the difference in the reflection angles, and the frequency estimate degrades. Estimation accuracy in this region could potentially be further improved with frequency estimates derived from second order statistics \cite{souden2009robust}, or by employing super-resolution techniques \cite{mamandipoor2016newtonized}.  However, at shorter ranges, more sophisticated frequency estimation should
be coupled with more detailed anisotropic reflection models, hence we leave this as an interesting topic for future work (e.g., on how to track a moving emitter in urban canyons).


\section{Algorithm for Source Seeking \label{sec:algorithm}}

In this section, we describe and justify the strategy for planning the UAV trajectory by using frequency measurements and show that the proposed algorithm will converge to the true source direction with no prior information on the source location. The purpose is to draw UAV to the vicinity of the emitter as quickly as possible. We set $d_{v}  \ll d$ as the required distance between UAV and the source at which we declare the tracking process successful. We assume constant speed through the trajectory of the UAV and use a feasible trajectory for the motion of UAV. The goal is to minimize flight time.

We assume prior knowledge of the emitted signal carrier frequency $f_c$, but not of the carrier frequency offset $f_o$, which also drifts over time. 
The pilot beacon for a given frequency measurement contains $N$ symbols, with symbol period $T_s$, so that the measurement interval for
frequency estimation is $T = N T_s$.  The pilot beacons are repeated with the period of $T_{slot}$.


For the $n$th received beacon,
frequency measurements are obtained by applying FFT to the $N$ complex baseband samples of (\ref{eq:channel}), and the peak frequency $\hat{\omega}_i$ $i \in  {1,\cdots,N_{FFT}}$ is refined by using a quadratic interpolation with adjacent samples:
\begin{equation} \label{eq:freq_est}
\tilde{\omega}_n = \hat{\omega}_i + \frac{\hat{\omega}_{i-1}-\hat{\omega}_{i+1}}{2 (\hat{\omega}_{i-1}+\hat{\omega}_{i+1} - 2 \hat{\omega}_i )} \frac{2 \pi}{T_s N_{FFT}}
\end{equation}
Thus, we obtain a noisy estimate of the sum of carrier frequency offset, Doppler frequency and the frequency drift. 
We model this, together with the bearing angle measured by the UAV sensors, as 
\begin{align} 
\tilde{\omega}_n &=  \omega_n + n_{\omega,n} \nonumber\\
\tilde{\phi}_n &= \phi_n + n_{\phi,n}. \label{measurement_model}
\end{align}
where $n_{\omega,n}$  and $n_{\phi,n}$ are frequency measurement and bearing measurement noises, modeled as zero mean independent Gaussian random variables with
variances $\sigma_{\omega,n}^2$ and $\sigma_{\phi,n}^2$, respectively.  The bearing measurement error variance $\sigma_{\phi,n}^2$ is assumed to be constant throughout the flight. However, the frequency measurement error variance $\sigma_{\omega,n}^2$ can vary: it increases as RSS drops during fades,
and as Doppler spread increases as the UAV approaches the emitter.

While we do not model the UAV dynamics, we will restrict the algorithm described in the next section to use trimming trajectories, which have the desirable property that the tracking error dynamics and kinematics is time invariant and for which there are well-developed trajectory tracking controllers \cite{kaminer1998trajectory, nelson2007vector, sujit2013evaluation}. The orientation tracking error about the desired trimming trajectory is included in error parameter $n_{\phi,n}$.

\subsection{Trajectory adaptation}

The source tracking algorithm can be divided into two stages, discussed in more detail below. In the first stage, the UAV gets a rough estimate of the direction of the source by doing a circular motion.
This stage is optional, and can be removed at the expense of some inefficiency in flight time.  The second stage involves piecewise linear trajectories
with perturbations of bearing, with change in Doppler providing feedback signal for continuous trajectory corrections.
 
\subsubsection*{Stage 1 - Circular motion for initial trajectory estimate}
\begin{figure}[h]
\centering
\includegraphics[width=0.8\columnwidth]{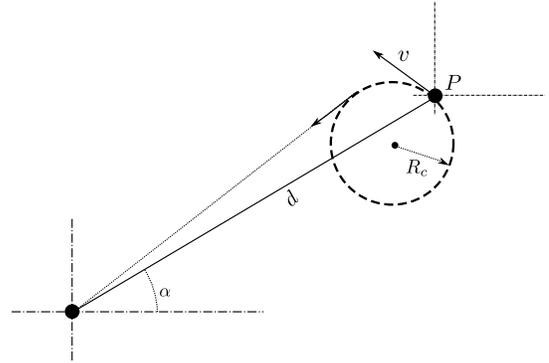}
\caption{Initial circular motion of the UAV}
\label{fig:circular}
\end{figure}

The UAV picks a random point at a distance $R_c$ and follows a circular trajectory, as shown in Figure \ref{fig:circular}, saving the frequency measurements $\tilde{\omega}_n$ with corresponding bearing measurements $\tilde{\phi}_n$. The largest frequency measurement corresponds to the maximum Doppler $f_{d,max}$ and bearing angle that corresponds to the desired direction is approximately $\pi + \alpha$ in an ideal setting. The smallest frequency measurements corresponds to the $-f_{d,max}$  at the direction of $\alpha$. 

The SNR is low when UAV is very distant, and multipath fading may occasionally result in large outliers in the frequency measurements $\tilde{\omega}_n$. 
We apply outlier rejection to the frequency measurements as 
\begin{equation} \label{eq:outlier_rejection}
\check{\omega}_n = 
\begin{cases}
    \tilde{\omega}_{n},& | \tilde{\omega}_{n} - \tilde{\omega}_{n-1}  | < 4 \pi \frac{v f_c}{c}\\
    \tilde{\omega}_{n-1},              & \text{otherwise}
\end{cases}
\end{equation}
and apply a moving average filter with length $15\ T_{slot}$. Then, the initial direction for UAV is determined by finding the direction of maximum frequency estimate as follows
\begin{align}  
i &=  \arg\!\max_n \check{\omega}_n \nonumber\\
\theta_0 & = \tilde{\phi}_i. \label{eq:step1}
\end{align}

\subsubsection*{Stage 2 - Continuous updates}
In this stage, the UAV derives information for feedback control of its trajectory in discrete time steps spanning $2M T_{slot}$ for each step.
If the estimated direction towards the emitter is $\theta_k$ from the previous direction, the UAV moves in the direction $\theta_k+\delta_k$ for a time interval with length $M T_{slot}$,
yielding frequency measurements $\{ \check{\omega}_m, m=1,...,M\}$, and then in the direction $\theta_k-\delta_k$ for the same duration, yielding measurements
$\{ \check{\omega}_m, m=M+1,...,2M\}$.
The difference between these two sets of frequency measurements is used to update $\theta_k$, as follows:
\begin{equation} \label{eq:step2}
\theta_{k+1} = \theta_k + \frac{1}{M} \big( \sum_{m=1}^{M}\check{\omega}_m - \sum_{m=M+1}^{2M}\check{\omega}_m \big) \frac{\delta_k}{2 \pi f_{d,max}}.
\end{equation}
Taking the difference in this fashion allows significant reduction of the effect of carrier frequency offset and drift, which vary slowly relative to 
the iteration step duration $2M T_{slot}$.  Additional robustness against measurement noise can be obtained by increasing the perturbation $\delta_k$, at the 
cost of increased travel distance.
 

\subsection{Analysis}

We now analyze the convergence of the algorithm in stage 2.
Straightforward trigonometry shows that the update step in equation (\ref{eq:step2}) with a constant $\delta_k=\delta$ can be written as 
\begin{equation}
\begin{aligned}
 \theta_{k+1} & = \theta_k + \left (  \cos(\theta_k - \theta_k^* + \delta) -\cos(\theta_k - \theta_k^* - \delta) \right ) \delta  \\
		      & =  \theta_k - 2 \sin(\theta_k - \theta_k^*)sin(\delta) \delta.
\end{aligned}
\end{equation}
Let the error term  $\tilde{\theta}_{k} = \theta_k - \theta_k^*$.  For $d \gg || p_{k+1} - p_k ||_2$ (range much larger than the distance between consecutive iterations), we have
$\theta_k^* \approx \theta_{k+1}^*$, which yields 
\begin{equation} \label{eq:error_Term}
\tilde{\theta}_{k+1} = \tilde{\theta}_{k} - 2 \sin(\tilde{\theta}_{k})sin(\delta) \delta 
\end{equation}
Intuitive insight is obtained for small $\tilde{\theta}_k$ and $\delta$ by using the approximation $\sin x \approx x$:
$\tilde{\theta}_{k+1} \approx \tilde{\theta}_{k} (1-2 \delta^2)$, corresponding to exponential decrease in estimation error.

For a rigorous proof of convergence, we pick $\tilde \theta_k^2$ as a Lyapunov function. From (\ref{eq:error_Term}), we obtain that the change in one time step is given by
%
$$
\tilde \theta_{k+1}^2-\tilde \theta_k^2 =  -4 \alpha\sin\tilde\theta_k(\tilde\theta_k -\alpha\sin\tilde\theta_k) 
$$
where $\alpha=\delta\sin\delta$.  Note that $\sin x (x - \alpha \sin x) > 0$ for $0 < |x| < \pi$ and $0 < \alpha < 1$.  Thus, $\tilde \theta_k^2$ is a strictly decreasing function for $\alpha < 1$, which provides a wide range of choices for $\delta$.


\begin{figure}[h]
\centering
\includegraphics[width=0.8\columnwidth]{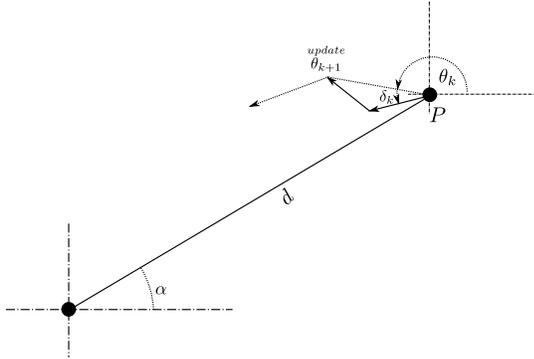}
\caption{Trajectory updates using direction perturbations $\pm \delta_k$}
\label{fig:step2}
\end{figure}

%
%

\section{Simulation Results}

\begin{figure}[h]
\centering
\includegraphics[width=0.8\columnwidth]{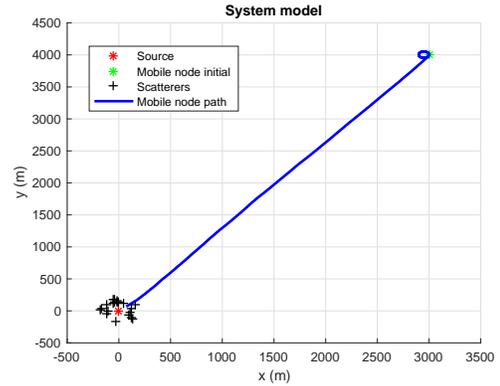}
\caption{Example trajectory (0 dB average initial SNR, initial distance 5 km).}
\label{fig:sim_results_trajectory}
\end{figure}

\begin{figure}[h]
\centering
\includegraphics[width=1\columnwidth]{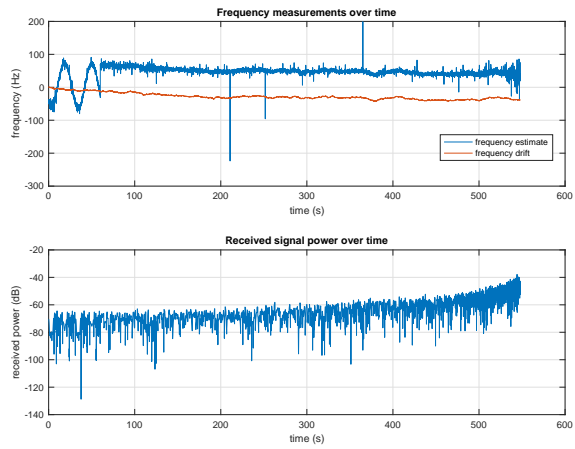}
\caption{Frequency measurements and RSS for the route in Figure \ref{fig:sim_results_trajectory}}
\label{fig:sim_results_freq_RSS}
\end{figure}

The simulation parameters are given in Table \ref{table:1}. We apply $N_{FFT} = 4096$ point FFT to $N=1000$ data chunks in every $T_{slot} = 50 ms$ for frequency estimation. The average received SNR at the initial distance of 5 km is set to 0 dB. Figure \ref{fig:sim_results_trajectory} shows an example UAV trajectory. Figure \ref{fig:sim_results_freq_RSS} shows the estimated frequency in the presence of multipath, CFO and frequency drift for that particular trajectory. Figure \ref{fig:sim_results_freq_RSS} also shows the received signal power profile through the trajectory and the spatial variations at the received power. 
We observe that the frequency estimation error increases as the UAV gets closer to the source due to increased Doppler spread.

\begin{table}[h!]
\begin{center}
\begin{tabular}{|c| |c|}
 \hline
 \multicolumn{2}{|c|}{Parameters} \\
 \hline
Parameter Symbol& Value \\
 \hline
d&5000 m  \\
\hline
R&200 m \\
\hline
$R_{in}$&100 m \\
\hline
$f_c$&2 GHz \\
\hline
v&10 m/s\\
\hline
$\sigma_n^2$&-70 dB\\
\hline
$T_{slot}$&50 ms\\ 
\hline
$T$&10 ms\\ 
 \hline
$T_s$&10 us\\ 
 \hline
 $N_{FFT}$& 4096\\ 
 \hline
 $\delta$&$10^o$\\ 
 \hline
  $d_v$&$200 m$\\ 
 \hline
 $M$&$20$\\ 
 \hline
\end{tabular}
\end{center}
\caption{Simulation parameters}
\label{table:1}
\end{table}

\begin{figure}[h]
\centering
\includegraphics[width=1\columnwidth]{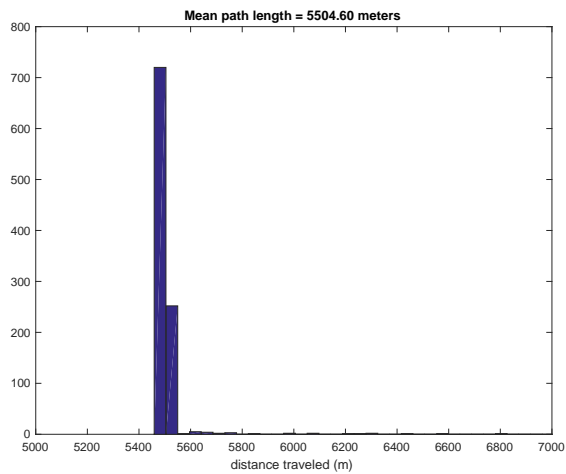}
\caption{Histogram of total distance traveled to get to within 200m of the source ($\textnormal{mean } \sim 5.5 \textnormal{km}$)}
\label{fig:hist}
\end{figure}

\begin{figure}[h]
\centering
\includegraphics[width=1\columnwidth]{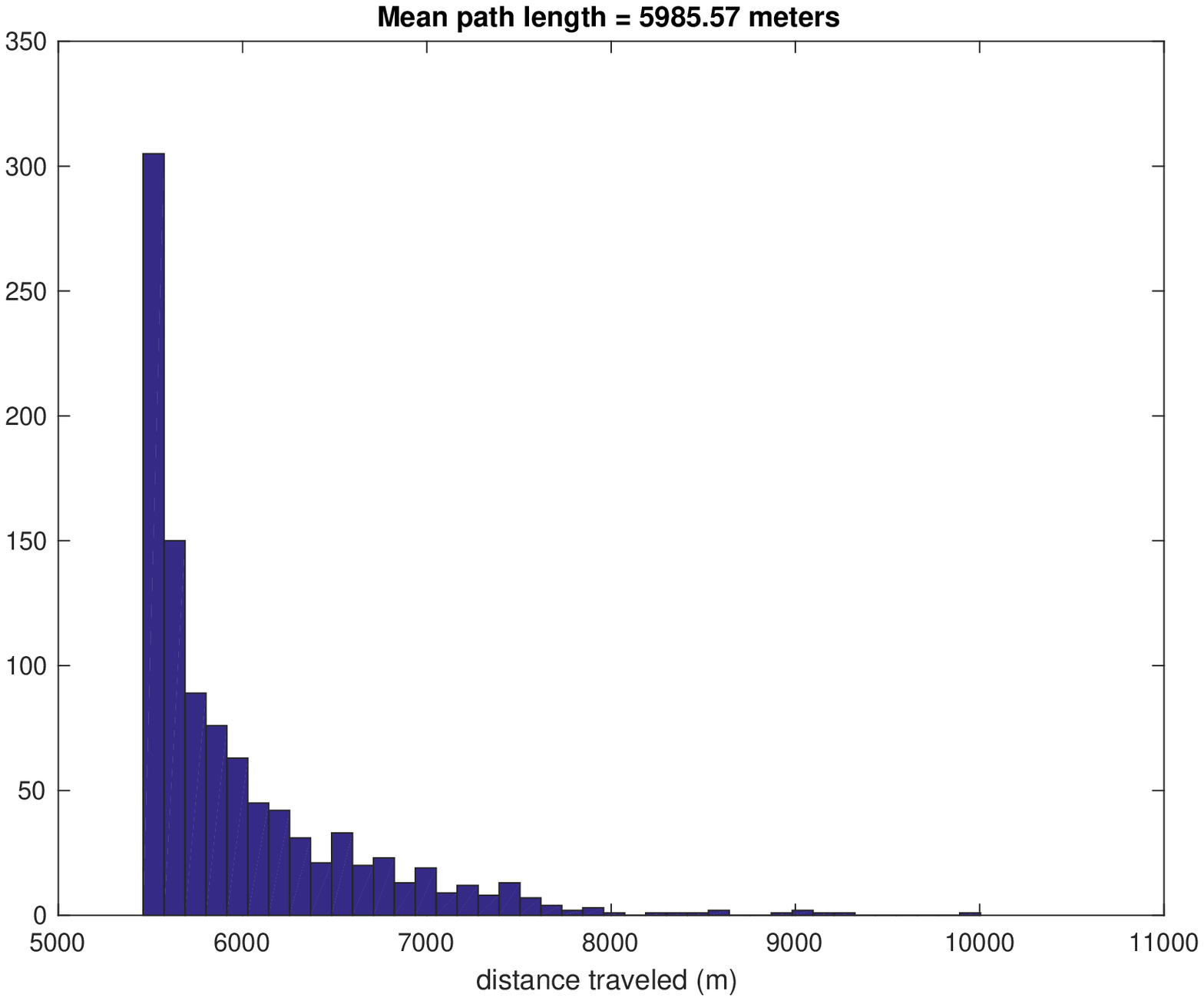}
\caption{Histogram of total distance traveled to get to within 200m of the source without stage 1 ($\textnormal{mean } \sim 6 \textnormal{km}$)}
\label{fig:hist2}
\end{figure}

Figure \ref{fig:hist} shows the histogram of the total distance traveled with Monte Carlo simulations of 1000 runs for the same scenario. The average distance traveled is 5.5 km, which is 1.15 times the shortest path to get within the desired distance of the target. This significantly outperforms the RSS based algorithm \cite{wadhwa2011following}, for which the average distance traveled
is about 3 times of the shortest path. The proposed algorithm works even if we discard stage 1 and use a random initial direction. Figure \ref{fig:hist2} shows the histogram of the total distance traveled with Monte Carlo simulations of 1000 runs with only Stage 2 of the algorithm. The average tracking distance is now 6 km,
which is 1.25 times the shortest path approach.

\section{Conclusions}

We have shown that a UAV with a single omnidirectional antenna can approach an RF source using only frequency and bearing measurements, in a manner
that is robust to multipath fading (via rejection of outliers in frequency measurements) and carrier frequency offset and drift (via averaging and differencing frequency measurements over relatively short intervals). 
Our analysis shows exponential convergence to the correct approach angle towards the source.   While the receiver operations required are more sophisticated than required for extracting RSS, the 
performance is far superior to that of a previously proposed RSS-based scheme.  
There are several interesting directions for future work, including
improved algorithms for determining the maximum Doppler, especially in the presence of Doppler spread; detailed modeling of the propagation environments
at shorter range, in order to understand the impact on both the frequency measurements and the trajectory updates (e.g., if the LoS is blocked, the UAV may follow a 
strong reflected path until it sees a LoS path again); and more detailed accounting of UAV dynamics.


\section*{Acknowledgments}
This research was supported by the National Science Foundation under grant CCF-1302114.


\bibliographystyle{IEEEtran}
\bibliography{references}

\begin{thebibliography}{10}
\providecommand{\url}[1]{#1}
\csname url@samestyle\endcsname
\providecommand{\newblock}{\relax}
\providecommand{\bibinfo}[2]{#2}
\providecommand{\BIBentrySTDinterwordspacing}{\spaceskip=0pt\relax}
\providecommand{\BIBentryALTinterwordstretchfactor}{4}
\providecommand{\BIBentryALTinterwordspacing}{\spaceskip=\fontdimen2\font plus
\BIBentryALTinterwordstretchfactor\fontdimen3\font minus
  \fontdimen4\font\relax}
\providecommand{\BIBforeignlanguage}[2]{{%
\expandafter\ifx\csname l@#1\endcsname\relax
\typeout{** WARNING: IEEEtran.bst: No hyphenation pattern has been}%
\typeout{** loaded for the language `#1'. Using the pattern for}%
\typeout{** the default language instead.}%
\else
\language=\csname l@#1\endcsname
\fi
#2}}
\providecommand{\BIBdecl}{\relax}
\BIBdecl

\bibitem{wadhwa2011following}
A.~Wadhwa, U.~Madhow, J.~Hespanha, and B.~M. Sadler, ``Following an {RF} trail
  to its source,'' in \emph{Communication, Control, and Computing (Allerton),
  2011 49th Annual Allerton Conference on}.\hskip 1em plus 0.5em minus
  0.4em\relax IEEE, 2011, pp. 580--587.

\bibitem{venkateswaran2013rf}
S.~Venkateswaran, J.~T. Isaacs, K.~Fregene, R.~Ratmansky, B.~M. Sadler, J.~P.
  Hespanha, and U.~Madhow, ``{RF} source-seeking by a micro aerial vehicle
  using rotation-based angle of arrival estimates,'' in \emph{American Control
  Conference (ACC), 2013}.\hskip 1em plus 0.5em minus 0.4em\relax IEEE, 2013,
  pp. 2581--2587.

\bibitem{becker1999passive}
K.~Becker, ``Passive localization of frequency-agile radars from angle and
  frequency measurements,'' \emph{IEEE Transactions on Aerospace and Electronic
  Systems}, vol.~35, no.~4, pp. 1129--1144, 1999.

\bibitem{nguyen2016single}
N.~H. Nguyen and K.~Do{\u{g}}an{\c{c}}ay, ``Single-platform passive emitter
  localization with bearing and doppler-shift measurements using pseudolinear
  estimation techniques,'' \emph{Signal Processing}, vol. 125, pp. 336--348,
  2016.

\bibitem{amar2008localization}
A.~Amar and A.~J. Weiss, ``Localization of narrowband radio emitters based on
  doppler frequency shifts,'' \emph{IEEE Transactions on Signal Processing},
  vol.~56, no.~11, pp. 5500--5508, 2008.

\bibitem{tahat2016look}
A.~Tahat, G.~Kaddoum, S.~Yousefi, S.~Valaee, and F.~Gagnon, ``A look at the
  recent wireless positioning techniques with a focus on algorithms for moving
  receivers,'' \emph{IEEE Access}, vol.~4, pp. 6652--6680, 2016.

\bibitem{witzgall2011Doppler}
H.~Witzgall, B.~Pinney, and M.~Tinston, ``Doppler geolocation with drifting
  carrier,'' in \emph{MILITARY COMMUNICATIONS CONFERENCE, 2011-MILCOM
  2011}.\hskip 1em plus 0.5em minus 0.4em\relax IEEE, 2011, pp. 193--198.

\bibitem{witzgall2015single}
H.~Witzgall, J.~Covington, and A.~Pierce, ``Single aircraft passive doppler
  location of radios,'' in \emph{Aerospace Conference, 2015 IEEE}.\hskip 1em
  plus 0.5em minus 0.4em\relax IEEE, 2015, pp. 1--8.

\bibitem{aragon2008antennas}
A.~Aragon-Zavala, \emph{Antennas and propagation for wireless communication
  systems}.\hskip 1em plus 0.5em minus 0.4em\relax John Wiley \& Sons, 2008.

\bibitem{baronkin2001cramer}
V.~M. Baronkin, Y.~V. Zakharov, and T.~C. Tozer, ``Cramer-rao lower bound for
  frequency estimation in multipath rayleigh fading channels,'' in
  \emph{Acoustics, Speech, and Signal Processing, 2001.
  Proceedings.(ICASSP'01). 2001 IEEE International Conference on},
  vol.~4.\hskip 1em plus 0.5em minus 0.4em\relax IEEE, 2001, pp. 2557--2560.

\bibitem{brown2010low}
D.~R. Brown~III, Y.~Liao, and N.~Fox, ``Low-complexity real-time single-tone
  phase and frequency estimation,'' \emph{IEEE Military Communication}, 2010.

\bibitem{souden2009robust}
M.~Souden, S.~Affes, J.~Benesty, and R.~Bahroun, ``Robust doppler spread
  estimation in the presence of a residual carrier frequency offset,''
  \emph{IEEE Transactions on Signal processing}, vol.~57, no.~10, pp.
  4148--4153, 2009.

\bibitem{mamandipoor2016newtonized}
B.~Mamandipoor, D.~Ramasamy, and U.~Madhow, ``Newtonized orthogonal matching
  pursuit: Frequency estimation over the continuum.'' \emph{IEEE Trans. Signal
  Processing}, vol.~64, no.~19, pp. 5066--5081, 2016.

\bibitem{kaminer1998trajectory}
I.~Kaminer, A.~Pascoal, E.~Hallberg, and C.~Silvestre, ``Trajectory tracking
  for autonomous vehicles: An integrated approach to guidance and control,''
  \emph{Journal of Guidance, Control, and Dynamics}, vol.~21, no.~1, pp.
  29--38, 1998.

\bibitem{nelson2007vector}
D.~R. Nelson, D.~B. Barber, T.~W. McLain, and R.~W. Beard, ``Vector field path
  following for miniature air vehicles,'' \emph{IEEE Transactions on Robotics},
  vol.~23, no.~3, pp. 519--529, 2007.

\bibitem{sujit2013evaluation}
P.~Sujit, S.~Saripalli, and J.~B. Sousa, ``An evaluation of {UAV} path
  following algorithms,'' in \emph{Control Conference (ECC), 2013
  European}.\hskip 1em plus 0.5em minus 0.4em\relax IEEE, 2013, pp. 3332--3337.

\end{thebibliography}

\end{document}